\documentclass{ws-ijmpa}

\begin{document}

\markboth{XU CAO, BING-SONG ZOU, and HU-SHAN XU} {Double pion
production in $NN$ and $\bar{N}N$ collisions}

%
\catchline{}{}{}{}{}
%

\title{Double pion production in $NN$ and $\bar{N}N$ collisions}

\author{XU CAO}

\address{Institute of Modern Physics, CAS, Lanzhou 730000, China\\
Graduate University of Chinese Academy of Sciences, Beijing 100049, China\\
Theoretical Physics Center for Science Facilities, CAS, Beijing
100049, China\\
caoxu@impcas.ac.cn}

\author{BING-SONG ZOU}

\address{Institute of High Energy Physics, CAS, Beijing 100049, China\\
Theoretical Physics Center for Science Facilities, CAS, Beijing
100049, China\\
zoubs@ihep.ac.cn}

\author{HU-SHAN XU}

\address{Institute of Modern Physics, CAS, Lanzhou 730000, China\\
Theoretical Physics Center for Science Facilities, CAS, Beijing
100049, China\\
hushan@impcas.ac.cn}

\maketitle

\begin{history}
\received{Day Month Year}
\revised{Day Month Year}
\end{history}

\begin{abstract}
With an effective Lagrangian approach, we give a full analysis on
the $NN \to NN\pi\pi$ and $\bar{N}N\to \bar{N}N\pi\pi$ reactions by
exploring the roles of various resonances with mass up to 1.72 GeV.
We find large contributions from $\Delta$, $N^*(1440)$,
$\Delta(1600)$ and $\Delta(1620)$ resonances. Our calculations also
indicate sizeable contributions from nucleon poles for the energies
close to the threshold. A good description to the existing data of
different isospin channels of $NN\to NN\pi\pi$ and $\bar{N}N\to
\bar{N}N\pi\pi$ for beam energies up to 2.2 GeV is reached. Our
results provide important implications to the ABC effect and
guildlines to the future experimental projects at COSY, HADES and
HIRFL-CSR. We point out that the \={P}ANDA at FAIR could be an
essential place for studying the properties of baryon resonances and
the data with baryon and anti-baryon in final states are worth
analyzing.

\end{abstract}

\ccode{PACS numbers: 13.75.Cs, 14.20.Gk, 25.75.Dw}

\section{Brief summary of current status}

Our understanding on the low energy strong interaction physics is
still unsatisfactory though quantum chromodynamics (QCD) has been
established as the standard theory of strong interaction for many
years. Because of its color confinement non-perturbative properties,
we have difficulties to calculate the properties of mesons and
baryons directly from QCD. Constituent quark models are developed to
explain the mesonic and baryonic spectrum and they are successful in
some aspects. But one of severe problems is that they predict more
excited states around 2 GeV than what have been observed
experimentally. As a result, a lot of efforts have been devoted to
the meson production in pion-, photo-, and electro-induced reactions
in order to reduce the uncertainties of extracted parameters of
resonances, deepen our knowledge on the structure of the resonances
and search for missing resonances.

In the past few years, the double pion production in nucleon-nucleon
collisions tends to be a fascinating field for studying resonances
properties and has been accurately measured at the facilities of
CELSIUS\cite{celsius}, COSY\cite{cosy}, KEK\cite{KEK}, and
PNPI-Gatchina\cite{Gatchina}.  In the Table~\ref{data}, we list all
the data measured after the year of 2000. The CELSIUS\cite{celsius}
and COSY\cite{cosy} Collaboration measured the differential cross
sections of $pp\to pp\pi^+\pi^-$ and $pp\to pp\pi^0\pi^0$ channels
for the beam energies from the threshold to 1.3 GeV. One of the
promising findings was that the $N^*(1440)$ contribution dominates
at the close-to-threshold region, as expected by the Valencia
model\cite{Valencia}. This provided us a good place to explore the
properties of $N^*(1440)$ whose theoretical interpretation is still
under controversial. However the Valencia model, though compatible
with the old bubble chamber and magnetic spectrometer
data\cite{lbbook}, overestimated the new data at the
close-to-threshold energies by several times. Another finding was
that the CELSIUS data of $\pi\pi$ invariant mass spectra for the
beam energies above 1.0 GeV demonstrated a single peak at the low
invariant mass while the model gave double hump structure which was
inconsistent to the data. The model also predicted a preferential
parallel emission of the two pion-meson, which was contrary to the
CELSIUS data. The third interesting finding was that there was a
level-off behavior around the beam energies of 1.0 GeV in the energy
dependence of the $pp\to pp\pi^0\pi^0$, which was thought to be the
result of interference between different contributions.  KEK
Collaboration made another progress and they measured the total
cross section of $pn\to pn\pi^+\pi^-$ and $pn \to pp\pi^-\pi^0$
channels which were important for our understanding on the
excitation of $N^*(1440)$.

\begin{table}[ph]
\tbl{The data of nucleon-nucleon collisions measured after the year
of 2000. Those with the data of differential cross sections are
marked by bold characters for the beam energies.}
{\begin{tabular}{@{}cc@{}} \toprule
Channel & Collaboration (Tp (MeV))  \\
\colrule
$pp\to pp\pi^+\pi^-$ \hphantom{00} & \hphantom{0}CELSIUS(\textbf{650, 680, 750, 775, 895, 1100, 1360}),\\
                                   & \hphantom{0}Gatchina(717, 818, 861, 900, 980), COSY(\textbf{750, 800}),\\
                                   & \hphantom{0}KEK(698, 780, 814, 908, 995, 1083, 1172)\\
$pp\to pp\pi^0\pi^0$ \hphantom{00} & \hphantom{0}CELSIUS(650, 725, 750, \textbf{775, 895, 1000, 1100, 1200, 1300}, 1360)\\
$pp\to nn\pi^+\pi^+$ \hphantom{00} & \hphantom{0}CELSIUS(800, \textbf{1100})\\
$pp\to pn\pi^+\pi^0$ \hphantom{00} & \hphantom{0}CELSIUS(725, 750, 775, 1100)\\
$pn\to pn\pi^+\pi^-$ \hphantom{00} & \hphantom{0}KEK(698, 780, 814, 908, 995, 1083, 1172)\\
$pn\to pp\pi^-\pi^0$ \hphantom{00} & \hphantom{0}KEK(698, 780, 814, 908, 995, 1083, 1172)\\ \botrule
\end{tabular} \label{data}}
\end{table}

On the other side, though the role of $N^*(1440)$ in nucleon-nucleon
collisions was firmly established, its contribution in $\bar{N}N\to
\bar{N}N\pi\pi$ reactions has never been considered.
JETSET\cite{JETSET} measured the $\bar{p}p \to \bar{p}p\pi^+\pi^-$
channel in order to search for narrow resonances decaying to
$\bar{p}p$ but they only included the double-$\Delta$ diagram in
their Monte-Carlo simulation. Moreover, it has never been explored
whether it was possible to extract the properties of other
resonances from antinucleon-nucleon collisions.

The one-pion exchange (OPE) models\cite{ope} of more than 45 years
ago only considered the double-$\Delta$ diagram and the Valencia
model of more than 10 years ago made a further step by including
also the $N^*(1440)$ resonance and the $\sigma$- and $\rho$-meson
exchange. With the new accurate data, it is very necessary to
perform a more comprehensive analysis including more resonances and
matching all the data of (anti)nucleon-nucleon collisions.

\section{Description of our model}

Our full model is demonstrated in Ref.~\refcite{caoprc}, so herein
we only give a brief description to the main features. The effective
Lagrangians for the resonances in our model are based on a Lorentz
covariant orbital-spin (L-S) scheme~\cite{zoucoupling,ouyang}. In
view of the overall system invariant mass about 2.8 GeV for Tp =2.2
GeV, we have checked contributions from all the well-established
$N^*$ and $\Delta^*$ resonances  below 1.72 GeV. The coupling
constants appearing in relevant resonances are determined by the
empirical partial decay width of the resonances taken from Particle
Data Group (PDG) book\cite{pdg2008}. Our calculated results show
that the $N^*(1440)$, $\Delta(1232)$, $\Delta^*(1600)$, and
$\Delta^*(1620)$ resonances play the relatively significant role in
the considered energies and other resonances give negligible
contributions~\cite{caoprc}.

In our model, the adjustable parameters are the cut-off values in
form factors at vertices and resonances. We include five types of
form factors: those at the meson-(anti)nucleon-(anti)nucleon
vertices, those at the meson-$N^*$($\Delta^*$)-(anti)nucleon
vertices, the Blatt-Weisskopf barrier factors at the
$N^*(1440)$-$\Delta$-$\pi$ vertex~\cite{zouform}, the form factor at
the $\sigma$-$\pi$-$\pi$ vertex, and the form factor of resonances
and nucleon poles. The $pp\to nn\pi^+\pi^+$ channel is very useful
to pin down these cut-off values in form factors of the relevant
$\Delta$ and $\Delta^*$ resonances because it has negligible $N^*$
contribution, and then we could easily determine the cut-off values
of $N^*$ resonances in other channels.

\section{Results and discussion}

\begin{figure}[pb]
\centerline{\psfig{file=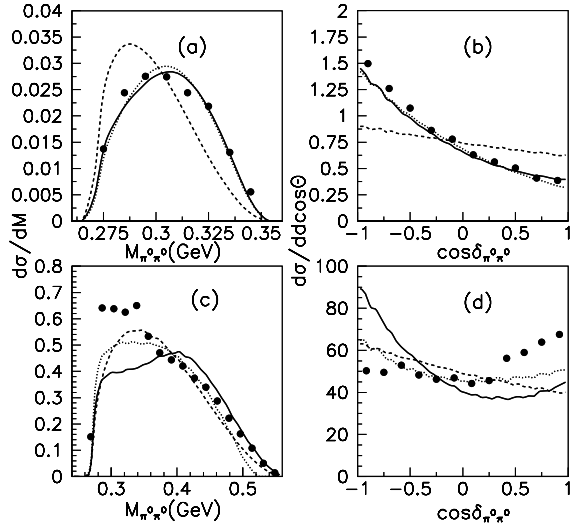,width=8cm}} \vspace*{8pt}
\caption{The invariant mass spectra and angular distributions of the
opening angle for $\pi^0\pi^0$ system\protect\cite{caoprc} in the
overall center-of-mass system for $pp\to pp\pi^0\pi^0$ at the beam
energies of 795 MeV (a, b) and 1300 MeV (c, d). The dashed and solid
curves correspond to the phase space and full model distributions ,
respectively. In (a, b), the dotted curves correspond to
$N^*(1440)\to N\sigma$ contributions. In (c, d), the dotted curves
correspond to double-$\Delta$ contributions. \label{pipidis}}
\end{figure}

\begin{figure}[pb]
\centerline{\psfig{file=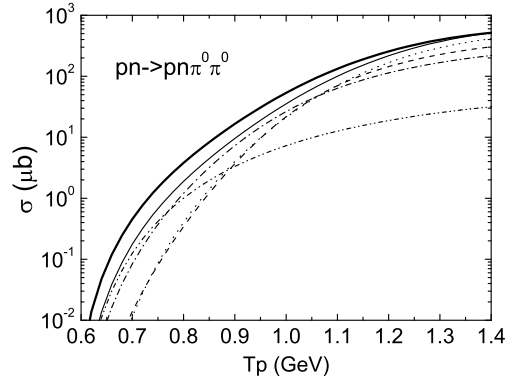,width=8cm}} \vspace*{8pt}
\caption{Total cross sections of $pn\to pn\pi^0\pi^0$ reaction. The
dotted, dash-dotted, dashed, dash-dot-dotted, solid and bold solid
curves correspond to contribution from double-$\Delta$, $N^*(1440)
\to N\sigma$, $N^*(1440) \to \Delta\pi$, nucleon poles, full model
results without and with FSI, respectively. \label{pppipi00}}
\end{figure}

If the double-$\Delta$ mechanism is dominant at high energies as in
the OPE model and Valencia model, the total cross section of $pp\to
pp\pi^0\pi^0$ should be a factor of about four larger than that of
$pp\to nn\pi^+\pi^+$ according to isospin coefficients. However, the
new exclusive measurements indicate an approximately equal value of
these two channels, which is consistent to the old bubble-chamber
data\cite{lbbook}. After calculating all the possible contributions,
our model got two important conclusions:
\begin{romanlist}[(ii)]
\item Our model reduces the relative branching ratio of $N^*(1440)\to\Delta\pi$
and assumes a smaller cutoff parameter for the $\pi N\Delta$
coupling so the relative contribution from the $N^*(1440)\to
N\sigma$ term increases significantly.
\item our model introduces significant contributions from
$\Delta\to N\pi\to N\pi\pi$ at energies near threshold and from
$\Delta^*(1600)$ and $\Delta^*(1620)$ at energies above 1.5 GeV.
This is more obvious in the $pp\to nn\pi^+\pi^+$ channel because the
isospin coefficients of these terms are much bigger than those in
other channels.
\end{romanlist}
As a result, our description of the total cross sections of all
isospin channels is considerably improved and our results agree well
with the new data of $NN$ collisions in the close-to-threshold
region. However, we find that it is difficult to interpret the
$\pi\pi$ invariant mass spectra for the beam energies above 1.0 GeV
and the level-off behavior around 1.0 GeV in the total cross section
of the $pp\to pp\pi^0\pi^0$ channel. In Fig.~\ref{pipidis}, we give
the invariant mass spectra and angular distribution of the opening
angle for $\pi^0\pi^0$ system in the overall center-of-mass system
in the $pp\to pp\pi^0\pi^0$ channel at the beam energies of 795 MeV
and 1300 MeV. At 795 MeV, which is close to the threshold, the
$N^*(1440)$ is dominant and the given spectra is consistent to the
data. At 1300 MeV the model fails to describe the data because the
$N^*(1440)\to\Delta\pi$, which is significant at high energies,
gives a double hump structure in the $\pi^0\pi^0$ invariant mass
spectrum. But if the $N^*(1440)$ contribution is negligible which is
caused by some kind of destructively interference in this energy
region, then the double-$\Delta$ diagram, which is dominant by
$\pi$-meson exchange in our model, describe the data
well\cite{celsius}. The situation is similar in the $pp\to
pp\pi^+\pi^-$ channel.

Our model provide essential hints to the ABC effect in $pn\to
d\pi^0\pi^0$ reaction\cite{abceffect}. In Fig~\ref{pppipi00} we show
the total cross section of $pn\to pn\pi^0\pi^0$ channel. It can be
seen that the $N^*(1440)\to N\sigma$ and $N^*(1440)\to\Delta\pi$ are
significant in all the considered energies. At the
close-to-threshold region, the nucleon poles are very important. So
it is suggested that the $N^*(1440)$ resonance and nucleon poles
should be carefully considered before we reach the right
conclusion\cite{Alvarezabc}.

The agreement between our model and the data of $\bar{N}N$
collisions is good\cite{ppbarcx}. It should be addressed that the
$N^*(1440)$ and other resonances should be included in the
Monte-Carlo simulation of the experimental analysis in order to get
the correct total cross sections from limited measured phase space,
which is overlooked by the previous measurements. We find that the
$\bar{p}n \to p\bar{n}\pi^-\pi^-$ reaction is useful to determine
the model parameters because like the $pp\to nn\pi^+\pi^+$ reaction,
the $N^*$ contribution is small. We also point out that other
channels in antinucleon-nucleon collisions could serve as a good
place for studying baryon resonances.

A lot of precise measurements in nucleon-nucleon collisions are
being carried out by CELSIUS and COSY collaborations in order to
study $N^*$ and $\Delta^*$ resonances. Besides, the HADES
collaboration has measured some channels of $NN$ collisions at the
beam energies of 1.25 GeV\cite{HADES} and their results are expected
to come soon, so this will constitute a good test of our model.
Recently, a cooler storage ring HIRFL-CSR, which can produce the
proton beam with similar beam energy range of COSY, has already been
successfully installed at Lanzhou. With the scheduled 4$\pi$
hadronic detector for complete measurement of diverse differential
cross sections, it will have a special advantage for studying
excited nucleon states through nucleon-nucleon collisions. We also
suggest that the double pion production in antinucleon-nucleon
collisions should be measured at PANDA (anti-Proton ANnihilation at
DArmstadt) at the GSI Facility of Antiproton and Ion Research
(FAIR), which could provide the antiproton beam of kinetic energy
ranging from 1 to 15 GeV with the luminosity of about
$10^{31}cm^{-2}s^{-1}$. We address that PANDA could play an
important role in the hadronic physics with a 4$\pi$ solid angle
detector with good particle identification for charged particles and
photons. The planned new experiments are anticipated to offer more
tools and information to help us understanding the low energy
physics better.

\section*{Acknowledgments}

Useful discussions with C. Wilkin, members of CELSIUS and HADES
Collaborations are gratefully acknowledged. This work was supported
by the National Natural Science Foundation of China (Grant Nos.
10635080, 10875133, 10821063, and 10925526).


\end{document}